\documentstyle[aps,preprint,psfig,epsf]{revtex}

\begin{document}

\title{Fermi Surface and gap parameter in high-$T_c$ superconductors:\\
the Stripe Quantum Critical Point scenario} 
\author{S. Caprara, C. Di Castro, M. Grilli,
A. Perali, M. Sulpizi} 

\address{Dipartimento di Fisica - Universit\`{a} di Roma ``La
Sapienza'' and Istituto Nazionale di Fisica della Materia, 
Unit\`a di Roma 1, P.le A. Moro, 2, - 00185 - Roma, Italy }

\date{\today} 
\maketitle 
\medskip

\begin{abstract}
We study the single-particle spectral properties of electrons coupled
to quasicritical charge and spin fluctuations close to a stripe-phase,
which is governed by a Quantum Critical Point near optimum doping.
We find that spectral weight is transferred from the quasiparticle peak
to incoherent dispersive features. As a consequence the distribution of 
low-laying spectral weight is modified with respect to the quasiparticle 
Fermi surface.
The interplay of charge and spin fluctuations reproduces features
of the observed Fermi surface, 
such as the asymmetric suppression of spectral weight
near the $M$ points of the Brillouin zone.   

Within the model, we also analyze the 
interplay between repulsive spin and attractive
charge fluctuations in determining the
symmetry and the peculiar momentum dependence of the 
superconducting gap parameter.
When both spin and charge fluctuations are coupled to the electrons,
we find $d_{x^2-y^2}$-wave gap symmetry in a wide range of parameter.
A crossover $d$- vs $s$-wave symmetry of the gap may occur when the
strength of charge fluctuations increases with respect to spin fluctuations.
\end{abstract}

\vskip 2pc 

\section{Introduction}
\label{intro}

A few theories for high-$T_c$ cuprate superconductors are based on the 
presence of a Quantum Critical Point ($QCP$) of some kind of instability 
in the phase diagram of these materials.

The different realizations of this scenario proposed so far involve
($i$) an antiferromagnetic ($AFM$)-$QCP$ \cite{sy,pines}, 
($ii$) a charge-transfer instability \cite{varma}, 
($ iii$) an ``as-yet unidentified'' $QCP$ regulating a first-order 
phase-transition between the $AFM$ state and the superconducting 
$SC$ state \cite{laughlin}, ($iv$) an incommensurate charge-density-wave 
($ICDW$)-$QCP$ \cite{cdg1,cdg2}.

The theories based on the $AFM$-$QCP$ \cite{sy,pines}
are motivated by the presence of an $AFM$ phase at low doping 
and by the observation of strong spin fluctuations at larger doping
\cite{neutronrm,neutronma,neutronpe,neutronmo}. 
However near and above optimum doping it is likely that charge 
degrees of freedom play a major role, whereas spin degrees of 
freedom {\sl follow} the charge dynamics, and are enslaved 
by the charge instability controlled by the $ICDW$-$QCP$\cite{cdg1,cdg2,jsc},
as suggested also by the experiments in $La_{2-x}(Nd Sr)_x Cu O_4$
\cite{tranq}.
The strong interplay between charge and spin degrees of freedom 
extends the spin fluctuations to a region far away from 
the $AFM$-$QCP$ and gives rise to a ``stripe phase'' which continuously
connects the onset of the charge instability ($ICDW$-$QCP$) at
high doping with the low-doping regime, where mobile holes, expelled
by the $AFM$ background, may form fluctuating stripes with marked one 
dimensional character \cite{zzz}.
Thus, we shall more properly refer to the $ICDW$-$QCP$ as the 
Stripe-$QCP$. The $AFM$-$QCP$ and the Stripe-$QCP$ are not 
alternative and they control the behaviour of the system in different 
regions of doping. On the other hand the existence of a Stripe-$QCP$ at 
optimum doping, where no other energy scale besides the temperature is
present in transport measurements \cite{T}, is the natural
explanation for the peculiar nature of this doping regime in the phase
diagram of superconducting copper oxides.
Indeed the critical fluctuations near a $QCP$ provide a singular
electron-electron ($e$-$e$) interaction which may account for
both the violation of the Fermi liquid ($FL$) behaviour in the metallic 
state and the high superconducting critical 
temperature \cite{cdg1,cdg2}.

In this paper we focus on the outcomes of the Stripe-$QCP$ scenario 
with respect to the single particle excitation spectra in the 
normal state and the symmetry of the gap parameter in the superconducting
state and we compare our results with the corresponding experiments, as 
discussed below.

The angle-resolved photoemission spectroscopy ($ARPES$), 
mainly performed on $Bi_2 Sr_2 Ca Cu_2 O_{8+\delta}$ ($Bi2212$), is providing 
new evidence for an anomalous metallic state in the under- and optimally
doped region of the phase diagram of high-$T_c$ cuprate superconductors.
The single-particle excitation spectra show features which cannot be 
described within the conventional Fermi liquid theory. The underdoped region
is characterized by the opening of a pseudogap around the $M$ points of
the Brillouin zone ($BZ$) below a crossover temperature $T^*$, 
which seems to evolve into the superconducting gap at the 
critical temperature $T_c$ \cite{ding,marshall,harris}. 
At the same time and in the same region of the $BZ$,
the quasiparticle peaks in the energy distribution curves have a rather
low spectral intensity and are very broad, even near the leading edge. 
At optimum doping, where the pseudogap disappears,
a strong suppression of low-laying spectral weight
around the $M$ points is still present \cite{marshall,Saini}.
Furthermore,
the recently developed angular scanning photoemission spectroscopy
provides a detailed picture of the $k$-space distribution of spectral
weight near the Fermi level and shows that the suppression around the 
$M$ points is asymmetric \cite{Saini}. The energy distribution curves are
characterized by the presence of dispersive peaks of incoherent nature 
besides the quasiparticle peak \cite{marshall}. 

In the superconducting state, the gap parameter has a peculiar doping
and temperature dependence. In the under- and optimally doped region 
of the phase diagram, $ARPES$ measurements on $Bi2212$ \cite{Campuzano}
and on $YBa_2Cu_3O_{7-\delta}$ ($Y123$) \cite{Shen93} have shown 
that the magnitude of the 
superconducting gap parameter is strongly momentum dependent in agreement
with a $d_{x^2-y^2}$-wave or a strongly anisotropic $s$-wave gap.
Several phase-sensitive measurements on $Y123$, involving
currents flowing within the $CuO_2$ planes \cite{Harl,Woll95,Tsuei},
together with temperature dependent penetration depth measurements 
\cite{Hardy}, provide evidence for primarily $d_{x^2-y^2}$-wave gap. 
On the other hand $c$-axis Josephson tunneling experiments, and in particular
the measurements where a conventional $s$-wave superconductor ($Pb$) is
deposited across a single twin boundary of $Y123$ 
twinned crystal, show that
the gap has a sizeable $s$-wave component \cite{twin}. 
 
In overdoped $Bi2212$ $ARPES$ experiments, out of the experimental error,
detect a finite value of the gap \cite{Kelley,Onellion}, 
hence the gap cannot have $d$-wave symmetry. 
The measured anisotropic ratio $r=\Delta_{max}/ \Delta_{min}$ 
is a decreasing function of the
doping, having $r\simeq 20$ near optimal doping, compatible with $d$-wave gap,
and $r\simeq 1.5$ at overdoping, compatible with anisotropic $s$-wave gap. 
In the same range of doping, the topology of the $FS$, including
the saddle points, does not change significantly \cite{Onellion}.
We take this as an indication that the strong doping dependence of the gap
parameter comes from a corresponding strong doping dependence of the
$e$-$e$ effective interaction.

\section{The normal state}

We study the single-particle excitation spectra of the metallic phase in 
the vicinity of the Stripe-$QCP$. The charge $(c)$ and (enslaved) spin ($s$)
fluctuations near the $QCP$, mediate an effective electron-electron
interaction, which in the random phase approximation can be written as

\begin{equation}
\Gamma_{eff}(q,\Omega)= \Gamma_c(q,\Omega)+\Gamma_s(q,\Omega)=
-\sum_{i=c,s}
\displaystyle{ V_i\over \kappa_i^2+\omega_i^2(q)
-{\rm i}\gamma_i \Omega },
\label{eff}
\end{equation}
where $\gamma_i$ is a damping coefficient and
$\omega_i^2(q)=2[2-\cos (q_x-q_x^i)-\cos (q_y-q_y^i)]$ 
is taken in the cos-like form to reproduce the $(q-q^i)^2$ behaviour
close to the wave-vector $q^i\equiv (q_x^i,q_y^i)$ 
of the critical charge and spin fluctuations, and to maintain the lattice
periodicity near the zone boundary. The distance from criticality is measured
by the inverse squared of the correlation length 
$\kappa_i ^2 \sim \max \left [a(\delta - \delta _c),bT\right ] $; 
we locate the Stripe-$QCP$
at a hole doping $\delta_c\simeq \delta_{optimal}$.
 The form (\ref{eff}) for the effective interaction, mediated by  
charge fluctuations, was found within a slave-boson 
approach to the Hubbard-Holstein model with long-range Coulomb
interaction, close to the charge instability \cite{cdg1,becca}. 
The same form, mediated by spin fluctuations, corresponds to
the phenomenological susceptibility proposed in Ref. \cite{millis} 
to fit NMR and neutron scattering experiments, in the strongly damped limit. 
The condition that charge fluctuations enslave spin fluctuations near the 
Stripe-$QCP$ is introduced, within our model, 
by requiring that the vanishing of $\kappa_c^2$
drives $\kappa_s^2$ to zero.

The bare electron dispersion law is taken in the form
\begin{equation}
\nonumber
\xi(k)=-2t(\cos k_x +\cos k_y)+4t'\cos k_x \cos k_y -\mu,
\end{equation}
where nearest-neighbour 
($t$) and next-to-nearest-neighbour ($t'$) hopping terms
are considered, to reproduce the main features of the band dispersion
and the $FS$ observed in high-$T_c$ materials. 
We choose $t=200$ meV, $t'=50$ meV, and $\mu=-180$ meV, 
corresponding to a hole doping $\delta \simeq 0.17$ 
with respect to half filling, as appropriate for $Bi2212$ at optimal doping.

Aiming to capture the essential aspects of the single-particle excitation
spectra, we calculate the electron self-energy within perturbation
theory, considering the first-order contribution  
$\Sigma(k,\varepsilon)= \Sigma_c(k,\varepsilon)+\Sigma_s(k,\varepsilon)$. 
The imaginary part of the self-energy is  
\begin{equation}
{\rm Im}\Sigma(k,\varepsilon)=\sum_{i=c,s}
V_i\gamma_i^{-1}\int_{BZ} 
{dk_x^\prime dk_y^\prime\over 4\pi^2}
\displaystyle{\left[f(\xi(k'))+b(\xi(k')-\varepsilon)\right]
[\varepsilon -\xi(k')] \over [\varepsilon -\xi(k')]^2 +\gamma_i^{-2}
[\kappa_i^2+\omega_i^2(k-k')]^2},
\label{imsigma}
\end{equation}
where the integral is extended over the $BZ$,
$f(\varepsilon)$ is the Fermi function and $b(\varepsilon)$ is the Bose
function. The real part of the self-energy ${\rm Re}\Sigma(k,\varepsilon)$
is obtained by a Kramers-Kr\" onig transformation of (\ref{imsigma}). 
To preserve the inversion symmetry $k\to -k$ in the resulting
quasiparticle spectra,
we symmetrize the self-energies $\Sigma_{c,s}$ with respect to $\pm q^{c,s}$.
For the sake of simplicity 
we assume that $q^s=(\pi,\pi)$, neglecting the possibility for a
discommensuration of the spin fluctuations in a (dynamical) stripe phase 
\cite{tranq}, which would introduce minor changes to the resulting
self-energy. The direction and the magnitude of the critical wave-vector 
$q^c$ are not universal. They depend on both the material \cite{mook} and 
the model \cite{goetz,zzz}.
To discuss the $ARPES$ experiments, which are mainly performed on
$Bi2212$ samples, we consider this material and we take  
$q^c=(0.4\pi,-0.4\pi)$ as it is suggested by the analysis 
of the experimental results \cite{Saini}.
The other parameters appearing in the effective interaction (\ref{eff})
are taken as $V_{c,s}=0.4$ eV, $\kappa_{c,s}^2= 0.01$ and
$\gamma_{c,s}=10$ eV$^{-1}$.
We focus on the quantum critical region near optimal 
doping, where the only energy scale is the temperature 
($\kappa_{c,s}^2\sim T$) and a strong violation of the Fermi-liquid behaviour
in the metallic phase is found.

To study the effect of the singular interaction (\ref{eff}) on the
single-particle properties in the metallic phase we calculate
the spectral density $A(k,\varepsilon)=\pi^{-1}|{\rm Im}\Sigma(k,\varepsilon)| 
/ \{ [\varepsilon-\xi(k)+\mu^{\prime}-{\rm Re}
\Sigma(k,\varepsilon)]^2+[{\rm Im}\Sigma
(k,\varepsilon)]^2 \}$. The chemical potential is 
self-consistently corrected by a term $\mu^{\prime}$ to keep 
the number of particles fixed.
To simulate the experimental conditions in $ARPES$ measurements
we introduce the convoluted spectral density
\begin{equation}
\nonumber
\tilde{A}_R(k,\varepsilon)=\int_{-\infty}^{+\infty}d\varepsilon'
A(k,\varepsilon')f(\varepsilon') E_{R}(\varepsilon'-\varepsilon),
\end{equation}
which takes care of the absence of occupied states above the Fermi energy,
through the Fermi function $f(\varepsilon)$, and of the experimental
energy resolution $R$, through a resolution function 
$E_{R}(\varepsilon)$. We take $E_R(\varepsilon)=
\exp(-\varepsilon^2/2R^2)/\sqrt{2\pi R^2}$ or
$=[\vartheta(\varepsilon+R)-\vartheta(\varepsilon-R)]/2R$
according to numerical convenience.

\begin{figure}[t]
\vskip 8.5cm
       \begin{picture}(8.0,5.0)
          \put(-6.0,-8.0){\epsfbox{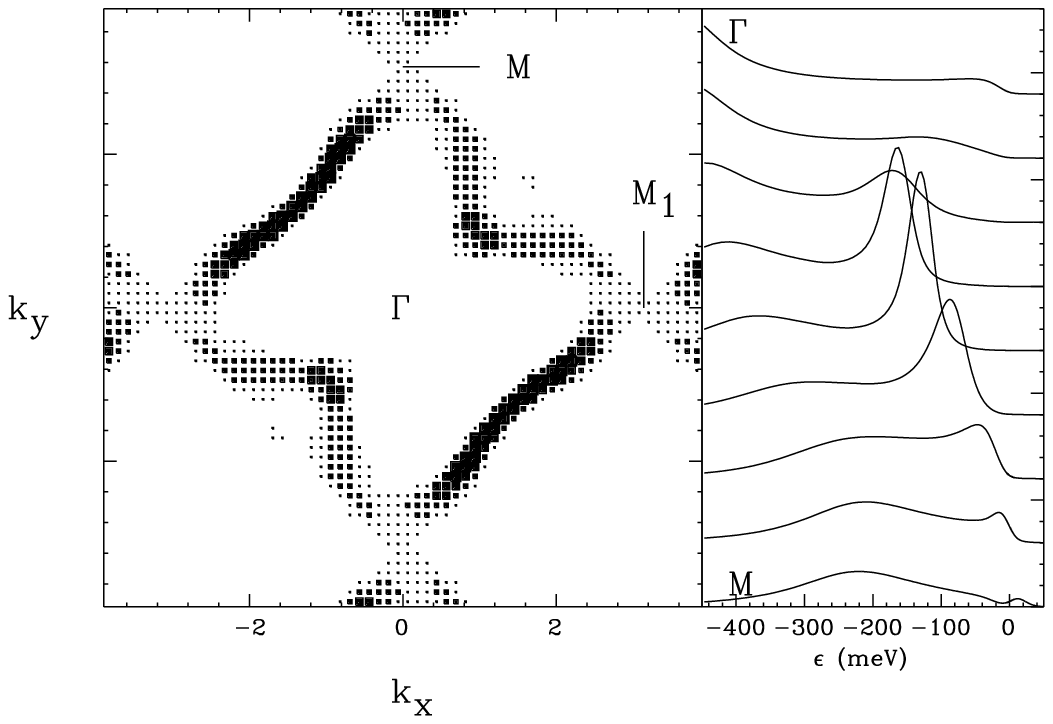}}
       \end{picture}
\caption{S. Caprara et al. }
\label{fs}
\end{figure}

The quasiparticle spectra are characterized by a coherent quasiparticle
peak at an energy $\varepsilon\simeq \xi(k)$ and by shadow resonances
at energies $\varepsilon\simeq \xi(k-q^i)$, produced by the interaction
with charge and spin fluctuations. The shadow peaks do not generally
correspond to new poles in the electron Green function and are essentially 
incoherent, although they {\sl follow} the dispersion of the shadow bands. 
Their intensity varies strongly with $k$ and increases as $\xi(k-q^i)$
approaches the value $\xi(k)$.
In particular at the hot spots, where $\xi(k)\simeq\xi(k-q^i)\simeq 0$ 
there is a suppression of the coherent spectral 
weight near the Fermi level due to the strong scattering.
In Fig. \ref{fs}, on the right panel, we plot the energy distribution curves
obtained along the direction $\Gamma M$ (from top to bottom). The 
quasiparticle peak moves from the left to the right towards the Fermi level
as the momentum is increased. It interferes with a shadow
resonance associated with $q^c$, which appears in the 
third to seventh curves. This resonance moves initially to the left
and then to the right, approaching the Fermi level. The effect of
spin fluctuations is visible only near the $M$ point, where a broad resonance
appears below the Fermi energy and is located at $-200$ meV at the $M$ point
in Fig. \ref{fs}.

We also study the $k$-distribution of low-laying spectral weight
$w_k={\tilde A}_R(k,\varepsilon=0)$. This distribution is more appropriate
than the standard definition of the $FS$ in terms of quasiparticles,
$\xi(k)+{\rm Re}\Sigma(k,\varepsilon=0)=0$, in the presence of incoherent
spectral weight near the Fermi level. Indeed,
the transfer of the spectral weight from the main $FS$ to the different 
branches of the shadow $FS$ at $\xi(k-q^i)\simeq 0$ produces features which
are characteristic of the interaction with charge and spin fluctuations and 
of their interplay. In particular
the symmetric suppression of spectral weight near the $M$ points
of the $BZ$, which would be due to spin fluctuations alone,
is modulated by charge fluctuations (Fig. \ref{fs}, left panel). 
This is also the case for the
(weak) hole pockets produced by spin fluctuations
around the points $(\pm \pi/2,\pm \pi/2)$. The interference with the branches 
of the shadow $FS$ due to charge fluctuations enhances these pockets 
around $\pm(\pi/2,\pi/2)$
and suppresses them around $\pm(\pi/2,-\pi/2)$ (Fig. \ref{fs}, left panel).
Experimental results on this issue are controversial. Strong shadow peaks
in the diagonal directions, giving rise to hole pockets in the $FS$,
have been reported in the literature \cite{Saini,larosa}, where
other experiments found only weak (or even absent) features \cite{marshall}.

We point out that, because of the transfer of spectral
weight to the shadow $FS$, the experimentally observed $FS$ may be rather
different from the quasiparticle $FS$. The observed evolution could, indeed,
be associated with the change in the distribution of the low-laying 
spectral weight, without any dramatic change in the topology of the 
quasiparticle $FS$, as was instead suggested in Ref. \cite{chubukov}.

We interpret the suppression of portions of the $FS$ as the onset of 
a pseudogap regime in the underdoped region.
We attribute the pseudogap formation to the strong pairing associated
with the singular $e$-$e$ effective interaction in the Cooper channel 
(see below, Eq. \ref{Cooper}). According to the scenario discussed in 
Ref. \cite{cdg2} the local gap formation should stabilize the system
against the true static charge ordering. The self-consistent 
interplay of the various effects should be considered within a more
refined model.

\section{Crossover between $d$- and $s$-wave gap parameter}

Recent $ARPES$ experiments indicate a crossover between 
$d$-wave and $s$-wave symmetry of the superconducting gap parameter 
driven by the increasing doping \cite{Kelley,Onellion}. 
In order to describe theoretically the change of the symmetry
of the gap parameter in the presence of an almost unchanged $FS$, 
the effective electron-electron ($e$-$e$) interaction has to be doping 
dependent.
This is the case for the effective interaction which arises 
near the Stripe-$QCP$ \cite{cdg1,cdg2,jsc,Perali}.
In a first approximation we consider the $e$-$e$ effective interaction
in the Cooper channel as a sum of the spin and charge fluctuation 
contributions coupled to fermions 

\begin{equation}
\label{Cooper}
\Gamma_{Cooper}(q,\Omega)=\Gamma_{c}(q,\Omega)-\Gamma_{s}(q,\Omega),
\end{equation}
where $\Gamma_{c}$ and $\Gamma_{s}$ are defined in formula (\ref{eff}),
which gives the effective $e$-$e$ interaction in the particle-hole channel.
Note that the spin fluctuations contribute with a different sign in the
$p$-$p$ and $p$-$h$ channels. 

In the under- and optimally doped region, 
the pairing is due to the interplay between critical charge 
fluctuations, providing in the small $q$ limit 
an attraction in both $d$-wave and $s$-wave
channels, and spin fluctuations providing attraction only in the $d$-wave
channel, being repulsive in the $s$-wave channel.
Indeed small $q$ interactions couple mainly nearby states in momentum
space which have the same sign of the $d_{x^2-y^2}$-wave symmetry factor 
($\cos k_x - \cos k_y$), preserving the attractive nature of the interaction.
We point out that the effective interaction mediated by charge fluctuations
includes, besides the small $q$ attraction, a residual repulsive term
at large $q$, as it was shown in Refs. \cite{cdg1,becca}. However in our 
calculation this repulsion is accounted for by the interaction mediated
by spin fluctuation. 
When both are considered, the $d_{x^2-y^2}$-wave gap parameter 
is enhanced by the cooperative effects of charge and spin fluctuations 
up to optimal doping \cite{jsc}.

In the overdoped region it is likely that
the spin fluctuations are strongly reduced, as
suggested by the decrease of the antiferromagnetic correlation 
length at high doping \cite{Pines95}. In this region of the phase diagram 
the charge fluctuations are dominant and the gap parameter may be 
$s$- or $d$-wave, depending on the strength of the residual repulsion
in the effective interaction.
This crossover appears as a possibility in our model and indeed may or 
may not occur depending on the values of the parameters of the effective
interaction as a function of doping.
A dominant role of spin degrees of freedom at low doping and of charge
degrees of freedom at higher doping in the pairing mechanism was also 
proposed in Ref. \cite{Sergio1} in the context of the three-band $t$-J-$V$ 
model.

As shown in Ref. \cite{Perali} 
the variation with doping of the charge instability vector $q^c$ is 
another possible mechanism to induce a symmetry crossover. 
When $q^c$ is less or of the same order of $2k_{Fy}$ (around $M$ points) 
the gap has a $d_{x^2-y^2}$-wave symmetry, 
since the interaction is characterized by small transferred momenta,
whereas for $q^c$ substantially greater 
than $2k_{Fy}$ the gap has an anisotropic $s$-wave structure.
In this last case the $d_{x^2-y^2}$-wave gap is strongly depressed because
the large $q$ interaction couples states with opposite sign of the 
corresponding symmetry factor, changing the attractive contribution into 
a repulsive one.
In the cuprates these two mechanisms could coexist.

In the overdoped region, where we assume that $\kappa _s ^2$ is sufficiently
large, the $q$ dependence of the effective interaction in the spin channel
introduces minor correction with respect to the dominant term 
$\Gamma _s (q=q^s,\Omega =0)$. The relevant contribution of $\Gamma_{s}$ to
the effective interaction (\ref{Cooper}) is given by

\begin{equation}
\nonumber
\Gamma_{s}(q,\Omega)\simeq\Gamma_{s}(q^s,0)
=-\frac{V_s}{\kappa_s^2}=-U.
\end{equation}
Thus, to simplify the discussion, we use in the following an
effective interaction  

\begin{equation}
\Gamma_{Cooper}(q,\Omega)=U+\Gamma_{c}(q,\Omega).
\label{coop}
\end{equation}
Within this approximation the contribution of the spin channel to
the effective interaction (\ref{Cooper}) has the same effects as
a residual repulsive interaction in the charge channel. 
In order to evaluate the symmetry and the momentum 
dependence of the superconducting gap,
we consider the static part of the effective interaction given by
$\Gamma(q)=\Gamma_{Cooper} (q,\Omega =0)$, and we solve the
$BCS$ equation for the gap parameter $\Delta(k)$. 
The numerical value of the gap, obtained  in the $BCS$ approach, is only
indicative because it is influenced by 
strong-coupling self-energy and non-adiabatic vertex corrections 
which have to be included 
in a complete generalized Eliashberg approach \cite{Grimaldi}. 
The $BCS$ equation is given by

\begin{equation}
\nonumber
\Delta(k) = -\frac{1}{N}\sum_{p_x,p_y}\Gamma (k-p)
\frac{\tanh\frac{\varepsilon(p)}{2T}}{2\varepsilon(p)}
\Delta(p),
\end{equation}
where $\varepsilon(p)=\sqrt{\xi^2(p)+\Delta(p)^2}$,
and the bare electron dispersion $\xi(p)$ is a tight binding fit
of the $Bi2212$ $FS$ measured by $ARPES$ at optimum doping \cite{Norman95}.
The $BCS$ equation is numerically solved taking advantage of the 
fast Fourier transform, as explained in Ref. \cite{Perali}.

Due to the qualitative character of the following discussion, we choose here
a direction for the characteristic wave-vector $q^c$ which emphasizes the 
$d$- vs $s$-wave crossover, and we take $q^c=(0,\frac{\pi}{4})$ along the 
$\Gamma M$ direction. 
We consider a small mass term $\kappa _c ^2 =0.1$, 
pushing the system into the Momentum Decoupling regime \cite{VPCP}.

\begin{figure}[!htb]
\protect
\centerline{\psfig{figure=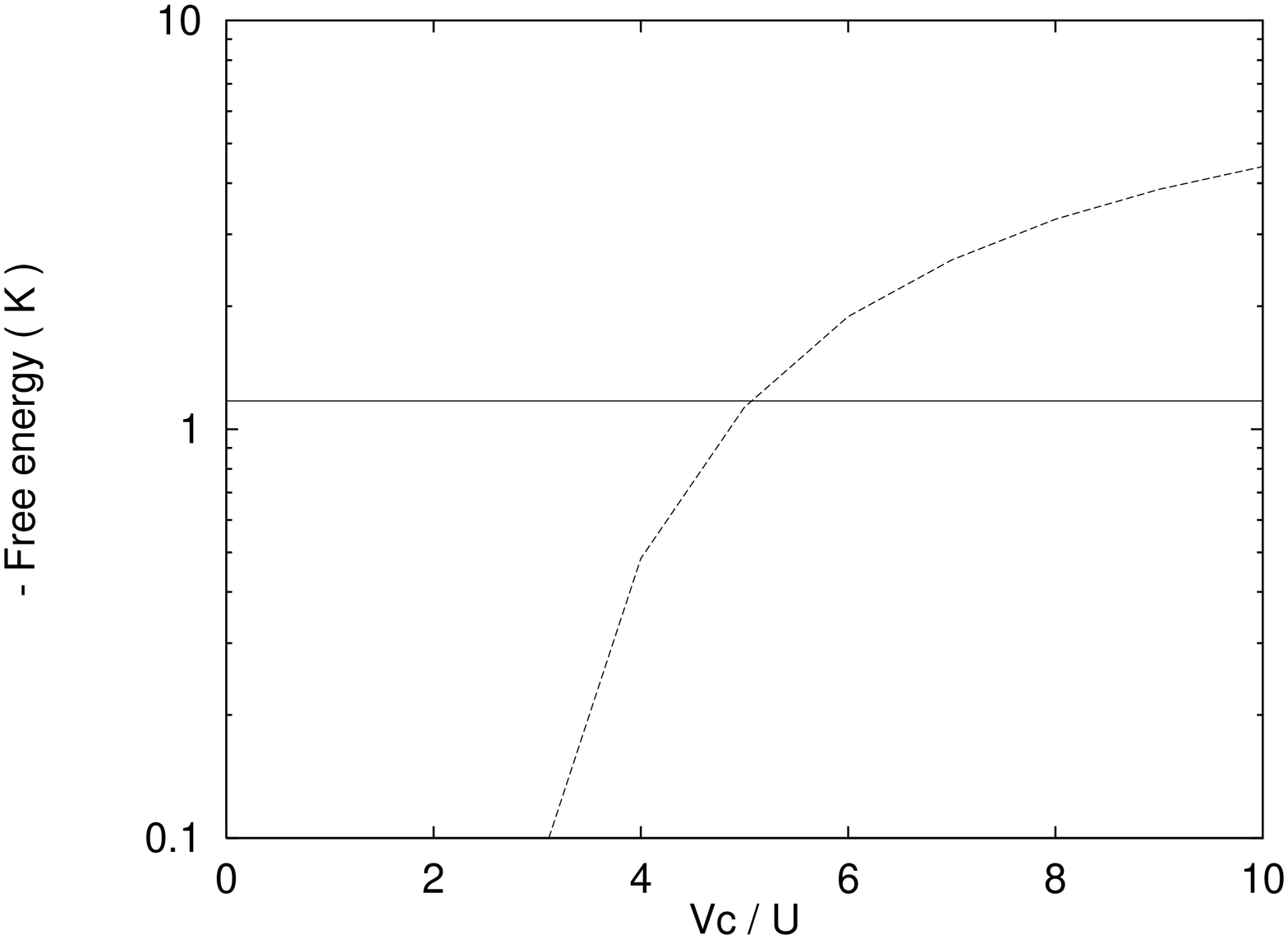,width=9cm,angle=-90}}
\caption{S. Caprara et al. }
\label{crossover}
\end{figure}

In Fig.\ref{crossover}
we report the condensation energy per particle as a function of
$\frac{V_{c}}{U}$ for fixed $V_c=0.4eV$ for both the 
$d$- and $s$-wave symmetry:
for $\frac{V_{c}}{U}<5$ (up to optimum doping) the
superconducting ground state has $d_{x^2-y^2}$-wave symmetry while for
$\frac{V_{c}}{U}>5$ 
(which should correspond to the overdoped case) the symmetry is $s$-wave. 

\begin{figure}[!htb]
\protect
\centerline{\psfig{figure=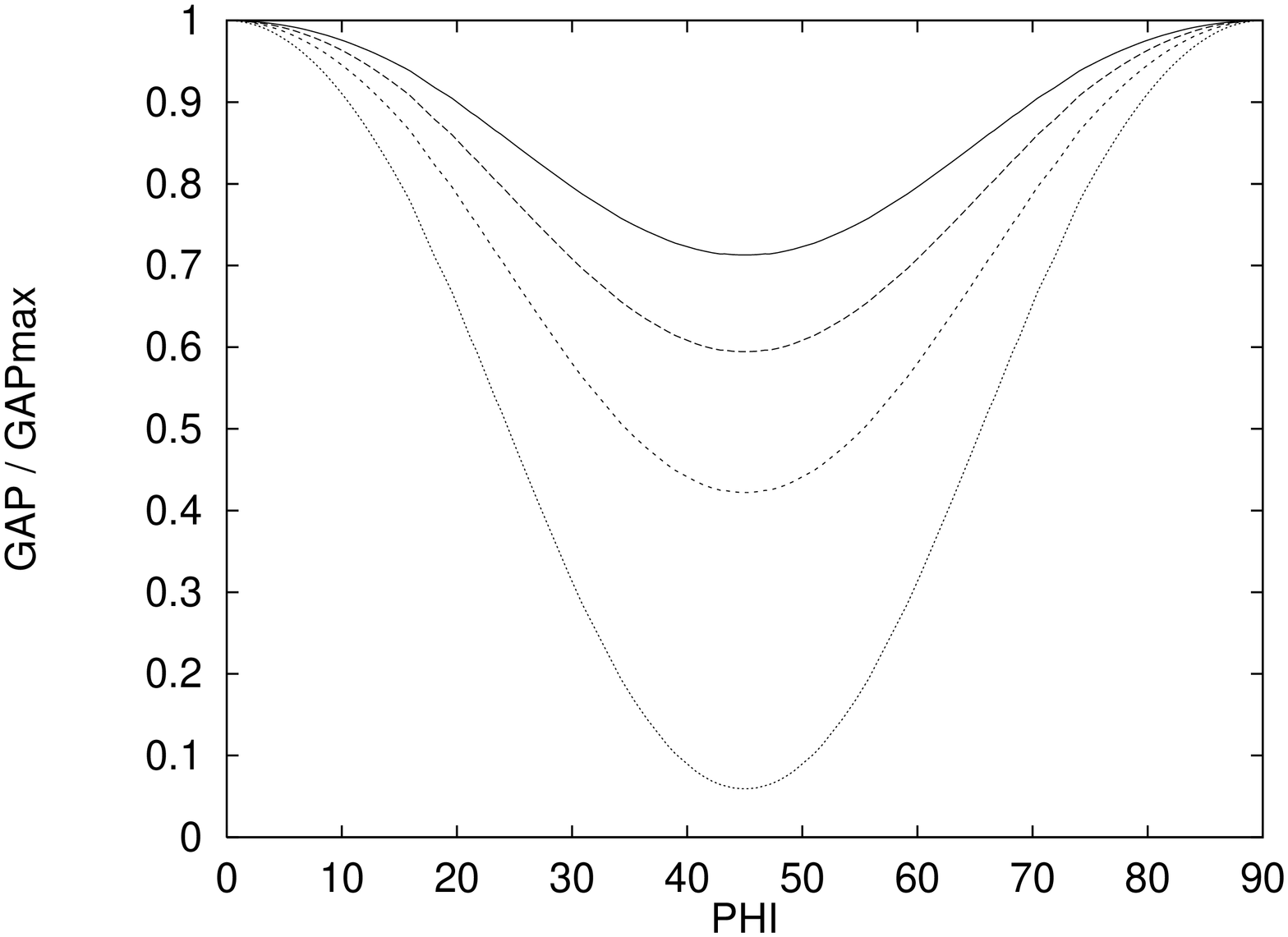,width=9cm,angle=-90}}
\caption{S. Caprara et al. }
\label{swave}
\end{figure}

The $s$-wave gap parameter, obtained for small values of $U$, 
$i.e.$ large doping, has an anisotropic momentum dependence along the $FS$
(Fig.\ref{swave}). The $s$-wave gap parameter, 
normalized to its maximum value $\Delta_{max}$, 
is plotted as a function of the angle $\phi$ which specifies the position 
of the Fermi momentum $k_F$ on the $FS$ with respect to the center
of the closed $FS$ for holes, which is located at the point $Y=(\pi,\pi)$.
The anisotropic ratio $r=\Delta_{max}/\Delta_{min}$
is a function of $U$, at fixed $\kappa _c^2 =0.1$ 
and $q^c=(0,\frac{\pi}{4})$.
We find $r=16.8$ for $U=0.12eV$, $r=2.4$ for $U=U^* =0.08eV$
(where $U^*$ is the crossover value between $d$- and $s$-wave symmetry) 
and $r=1.68$ for $U=0.04eV$ giving a decreasing
anisotropy for decreasing $U$, 
which corresponds to increasing doping, in agreement 
with the trend observed in $ARPES$ experiments \cite{Kelley,Onellion}.
Notice, however, that from Fig.\ref{crossover}, the transition from $d$- to 
$s$-wave superconductivity occurs at a $U/V_c\simeq 0.2$ ratio, where the
anisotropy ratio $r\simeq 2.4$ is rather small. Therefore, at least for the 
(reasonable) parameters considered here, the $d$- to $s$-wave transition would 
be characterized by a rather abrupt decrease of anisotropy. This expectation
could be tested experimentally by a detailed $ARPES$ analysis at closely spaced 
fillings.

\section{Conclusions}
\label{concl}

In this paper we have   
briefly described the Stripe-$QCP$ scenario and reported some of its
consequences in the normal and superconducting states. 
Coming from the high doping regime,
the occurrence of a charge-ordering instability towards a stripe phase, 
which is masked by the onset of a superconducting phase, 
provides the mechanism which controls the physics of the cuprates. 
It gives rise to the non-Fermi liquid properties of the normal phase, 
to some features found in $ARPES$ experiments, and to the strong 
pairing interaction leading to $d$-wave superconductivity.
 
The main additional results of this paper concern
the shape of the Fermi surface in the proximity of the Stripe-$QCP$ 
and the symmetry of the superconducting gap, in comparison 
with the $ARPES$ experiments on optimally \cite{Saini} 
and over- doped $Bi2212$ \cite{Kelley,Onellion}. 

By considering in our model a diagonal charge instability 
vector $q^c=(0.4\pi; -0.4\pi)$, as suggested by the experiments \cite{Saini},
we reproduce the asymmetrical spectral weight suppression around 
the $M$ points observed in the normal state.
We also describe the occurrence of shadow resonances in the energy
distribution curves, which appear in addition to the 
quasiparticle peaks as a result of the interaction
of the fermions with charge and spin quasicritical fluctuations.

In the superconducting state we obtain a $d_{x^2-y^2}$-wave gap parameter
in a wide range of parameters and we show that, within our scenario,
it is possible to produce a $d$- $vs$ $s$-wave
gap symmetry crossover, and a reduction of the anisotropy of the $s$-wave
gap with increasing doping, depending on the evolution of the parameters
entering the effective interaction, as a function of doping.
The main mechanisms which stabilize the anisotropic $s$-wave gap parameter,
leading to the symmetry crossover, are the reduction of the residual 
repulsion and/or the increase of the charge instability 
vector $q^c$ upon doping. 
A similar crossover
was recently reported in $ARPES$ experiments on 
overdoped $Bi2212$ \cite{Kelley,Onellion}.
Further experimental investigations are required to confirm a doping 
dependent crossover and to have a feedback on the theory.  

{\bf Acknowledgements:}
Part of this work was carried out with the financial 
support of the INFM, PRA 1996. The authors would like to thank 
Prof. C. Castellani for many useful discussions and suggestions.

\newpage

Fig. 1 LEFT: $k$-space distribution of low-laying spectral weight. 
The relative intensity decreases by a factor of two as the size of the black 
square is reduced. The energy resolution is taken $R=25$ meV, and the 
temperature is $T=30$ meV to compare with the results in Ref. \cite{Saini}. 
The values 
of the other parameters are given in the text. RIGHT: Energy distribution 
curves along the $\Gamma M$ direction for uniformly increased wave-vector 
$k$. The resolution is taken $R=10$ meV, and the temperature is $T=10$ meV, to 
compare with the results in Ref. \cite{marshall}.

Fig. 2 Condensation energy per particle for $V_{c}=0.4eV$
and increasing $U$: $d$-wave (full line); $s$-wave (dashed line).
The $d$-wave free energy is independent of $U$, because the 
momentum-independent repulsive contribution vanishes in the $d$-wave channel.
According to our discussion an increasing $V_c/U$ corresponds to increasing
doping. 

Fig. 3 Normalized $s$-wave gap for vertical direction 
with $q^c_y=\frac{\pi}{4}$
and decreasing values of $\frac{U}{V_c}=0.3$ (dotted); $0.2$ (short dashed);
$0.1$ (long dashed); $0.0$ (full line).

\newpage

\end{document}